\documentclass[twocolumn]{svjour3}
\smartqed
\usepackage{graphicx}
\usepackage{siunitx}
\usepackage{braket}
\usepackage{lineno}
\usepackage{ulem} 
\usepackage{lmodern}
\usepackage[utf8x]{inputenc}
\usepackage{todonotes}
\usepackage{hyperref}
\journalname{myjournal}
\begin{document}
\title{An extended-cavity diode laser at \SI{497}{nm} for laser cooling and trapping of neutral strontium}

\author{Vladimir Schkolnik$^{\dagger}$ \and Oliver Fartmann \and Markus Krutzik 
}


\institute{Vladimir Schkolnik   \and Oliver Fartmann \and Markus Krutzik \at Humboldt-Universit\"at zu Berlin, Newtonstr. 15, 12489 Berlin, Germany\\    \email{vladimir.schkolnik@physik.hu-berlin.de}          
             \\\
}
\date{}
\maketitle
\begin{abstract}

We present the first extended-cavity diode laser in Littrow configuration operating in the cyan wavelength range around \SI{497}{nm}. The gallium-nitride based diode laser features a free-space output with up to \SI{60}{mW}, operates in a single frequency mode, is tunable over a range of more than \SI{8}{nm} and has a Lorentzian linewidth of less than \SI{90}{kHz}. A detailed characterization of the tuning capabilities of the diode laser and its emission spectrum is provided. This compact, simple and low cost laser source replaces more complex systems based on frequency doubling and therefore simplifies the development of future compact and mobile optical clocks based on neutral strontium. Applications include efficient repumping of strontium atoms from the \mbox{$5s5p$  $^3P_2$} state and the \SI{9.8}{MHz} broad $5s5p$  $^3P_2  \rightarrow 5s5d$  $^3D_3 $ transition might be of interest for sub-Doppler cooling.

\end{abstract}


\section{Introduction}
\label{sec:intro}

 Nowadays, optical atomic clocks reach fractional uncertainties at the \SI{2 e-18}{} level \cite{Nicholson2015} and relative instabilities at the \SI{e-17}{} level at one second of averaging time \cite{Sanner2018}. Due to their high accuracy and stability, optical clocks are used in various applications in fundamental research, pushing the limits in tests of the time variation of fundamental constants \cite{Safronova2018b}, tests of general relativity \cite{Dzuba2017}, search for dark matter \cite{Roberts2017,Kalaydzhyan2017} and are envisioned as detectors for gravitational waves \cite{Kolkowitz2016}.  Their high precision makes optical clocks also  important tools for applications like navigation \cite{Schiller2012} and geodesy \cite{Mehlstaubler2018,Flury2016}.

Currently the best performing optical clocks are based on neutral atoms trapped in a magic wavelength optical lattice \cite{Ludlow2015}.  In particular, clocks based on neutral strontium have become widely used over the world in the past, operating uninterrupted over multiple weeks in laboratories and recently started to even perform outside the lab \cite{Grotti2018}.
 
\begin{figure}[b]
\centering
\includegraphics[width=0.95\linewidth]{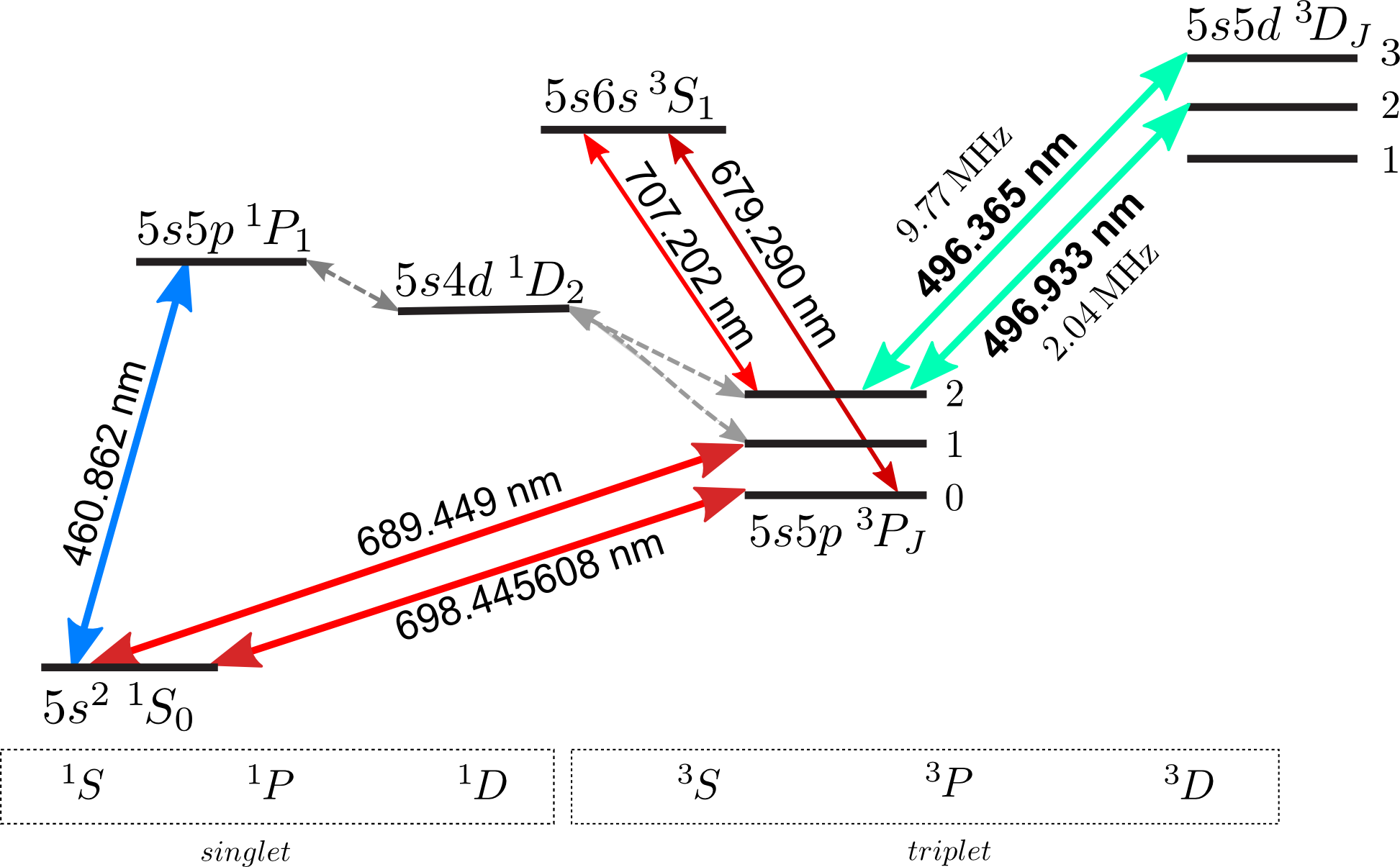}
\caption{The level scheme for the bosonic strontium isotope $^{88}$Sr with  transitions used for laser cooling, repumping, clock operation and detection and their vacuum wavelengths (linewidth given for the cyan transitions).}
\label{fig:Sr_scheme}
\end{figure}

However, the general scheme for laser cooling and trapping of neutral strontium and other alkaline earth atoms is more complex compared to the one used for alkali atoms \cite{Metcalf2003}. The cooling process starts with a first cooling stage utilizing the broad $^1S_0 \rightarrow$  $^1P_1$ transition at \SI{460.9}{nm}, from where atoms can decay towards the meta-stable $^3P_2$ state with a branching of roughly 1 in 50,000 (see Figure \ref{fig:Sr_scheme}),  which limits the MOT lifetime to a few tens of milliseconds \cite{Stellmer2014}. To increase the number of atoms in the MOT several repumping schemes have been employed. Due to the availability of laser sources the majority of strontium lattice clocks use two repump lasers addressing the $^3P_2 \rightarrow$  $^3S_1$ transition at \SI{707}{nm} and the $^3P_0 \rightarrow$  $^3S_1$ transition at \SI{679}{nm}.
Alternatively, only a single repump laser addressing the $^3P_2 \rightarrow$  $^3D_2$ transition at \SI{496.93}{nm} may be used \cite{Moriya2018,Poli2005}. Until now the generation of light at this wavelength relied on second harmonic generation (SHG) from an infrared laser due to the lack of laser diodes directly operating in this range. This results in complex setups with low optical output power but high power consumption that are undesirable for mobile applications.

Newly available gallium-nitride (GaN) laser diodes operating around \SI{495}{nm} pave the way to build extended-cavity diode lasers (ECDL) at \SI{497}{nm}. This allows for more compact and energy efficient laser sources for laser cooling and trapping of strontium.  Optical powers of up to \SI{100}{mW} open up the possibility of sub-Doppler cooling on the almost closed cycling transition $^3P_2 \rightarrow$  $^3D_3$ \cite{Stellmer2014,Hayakawa2018} as already successfully demonstrated with the alkaline earth metal calcium \cite{Gruunert2002}. This could lead to temperatures in the $\mu$K regime, sufficient for effectively loading atoms into the magic wavelength lattice and therefore allows to replace the usual laser system at \SI{689}{nm} that requires prestabilization to an optical cavity.

Here we present the first laser source at \SI{497}{nm} based on an extended-cavity diode laser in Littrow configuration. After presenting the characteristics of the used laser diodes in Chapter 2.1 we present the ECDL design in Chapter 2.2, followed by the characterization of its tuning and spectral characteristics in Chapter 2.3 and 2.4, respectively. We conclude with an outlook on future applications in Chapter 3.

\section{Extended-cavity diode laser at \SI{497}{nm}}
\label{sec:setup}

\subsection{Laser diodes}
\label{subsec:diodes}

In the past years GaN diodes covering the spectrum from \SI{370}{nm} to \SI{530}{nm} have become commercially available. For applications using neutral strontium atoms, diodes operating around the \SI{460}{nm} transition reach output powers of up to \SI{100}{mW} (Nichia corporation, \linebreak  NDB4216),  and with amplifier chips multiple thereof should be available in the near future. Diodes based on InAlGaN multiple quantum wells for the wavelength region around \SI{495}{nm} and \SI{500}{nm}, tunable to the $^3P_2  \rightarrow$ $^3D_2 $ and $^3P_2 \rightarrow$ $^3D_3 $ transition in neutral strontium at \SI{496.36}{nm} and \SI{496.93}{nm}, respectively, were recently introduced from Sharp (GH04955A2G) \cite{SharpGH04955A2G}. According to the data sheet the emission wavelength of these non AR-coated diodes spans from \SI{480}{nm} to \SI{495}{nm}. For our purposes we use diodes from a selected batch with a peak wavelength of \SI{495}{nm}.

\subsection{ECDL design}

\begin{figure}[htb]
\centering
\includegraphics[width=0.95\linewidth]{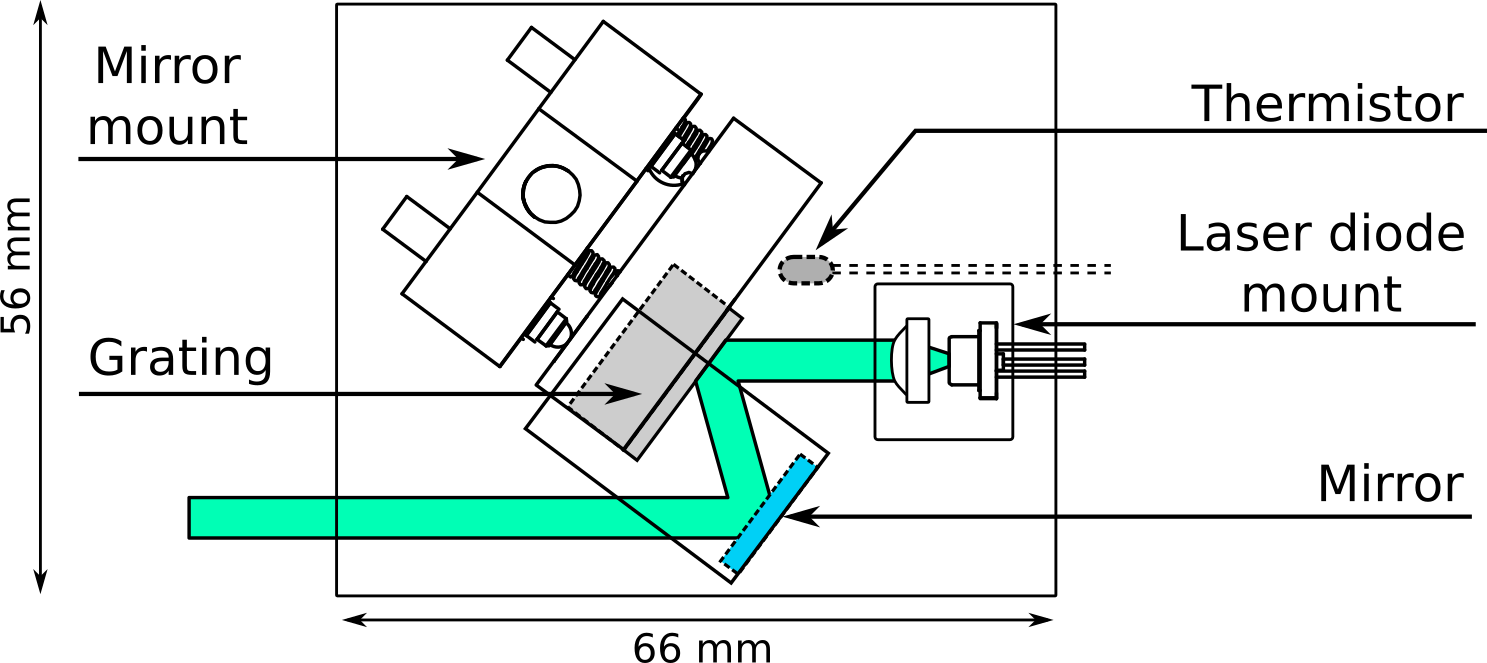}
\caption{Top-view schematic of the extended-cavity diode laser.
The Peltier element underneath the base plate (outer frame in the picture) is not shown.}
\label{fig:design_ECDL}
\end{figure}

The design of our diode laser is shown in Figure \ref{fig:design_ECDL}. It consists of an extended-cavity diode laser in Littrow configuration \cite{Ricci1995} and a mirror for angle compensation \cite{Hawthorn2001}.  Light emitted from the laser diode is collimated using an aspheric lens with a focal length $f = \SI{4.02}{mm}$. The collimated light from the diode illuminates a holographic reflective grating (Thorlabs, GH13-24V) and the $-1^{\text{st}}$ order is being reflected back into the laser diode providing the optical feedback. The reflected $0^{\text{th}}$ order is used as the output. The grating period $d$ of \SI{2400}{l/mm} results in a diffraction angle of \SI{36.5}{\degree}. The diffraction efficiency in the $-1^{\text{st}}$ order is about 38 \% and about 44 \% are reflected into the $0^{\text{th}}$ order.

The laser diode is mounted in a press fit aluminum mount inspired by the laser pointer community. This ensures good thermal contact between the diode and the rest of the assembly. The grating is embedded into a mirror mount (Radiant Dyes, MNI-H-U-2-3000) to reduce the cavity length and thus to increase the stability. Both the mirror mount and the diode mount are integrated onto a small aluminium base plate. The temperature is stabilized using a digital TEC controller (Meerstetter Engineering, TEC-1091) by means of a Peltier element and an NTC thermistor that is positioned in the base plate between the diode and the grating, inspired by \cite{Saliba2009b}. The diode mount and the mirror mount are fixed with nylon screws and finally glued to the base plate. A mirror is mounted on a small lever parallel to the grating to compensate for the angle change of the beam due to wavelength tuning with the grating angle \cite{Hawthorn2001}. Our diode laser current source is based on the Libbrecht-Hall design \cite{Libbrecht1993}. \SI{86}{\percent} of laser emission passes an optical isolator (LINOS, FI-488-3SC) and is coupled into a polarization-maintaining optical fiber with an efficiency of $\approx$\,\SI{55}{\percent} for optical spectrum analysis.

\subsection{Power and tuning characteristics}

\begin{figure}[htb]
\centering
\includegraphics[width=0.95\linewidth]{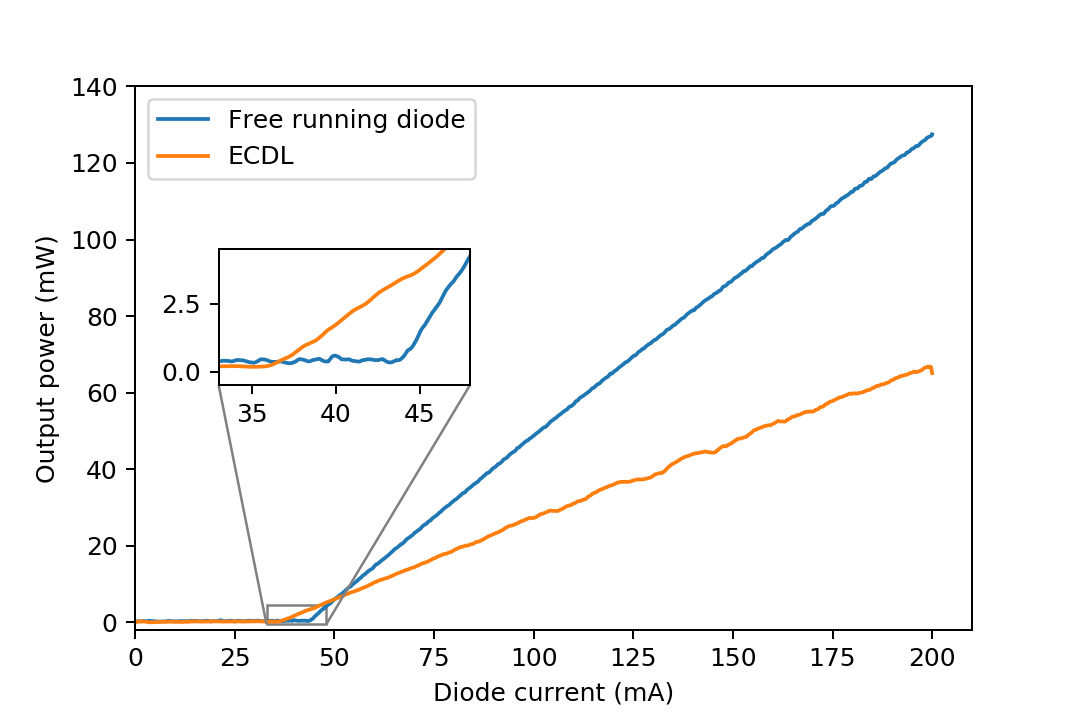}
\caption{The output power of the bare laser diode (blue) and of the ECDL with optical feedback from the grating (orange) as a function of the injection current at \SI{25}{\celsius}. The inset shows the reduced laser threshold of the ECDL. }
\label{fig:PI}
\end{figure}

We measured the output power of the bare laser diode and of the assembled ECDL both at a temperature of \SI{25}{\celsius}. The power was measured before the optical isolator. The results are shown in Figure \ref{fig:PI}. According to the data sheet, the absolute maximum injecting current is \SI{135}{mA} with a typical output power of \SI{55}{mW} at \SI{105}{mA}. In our measurement, we operated the laser diode at injection currents up to \SI{200}{mA}. As can be seen from Figure \ref{fig:PI}, the output power of the free running diode reaches \SI{127}{mW} with only a slight deviation from a linear current/power relationship of \SI{5.9}{\percent}. Therefore, an output power of almost 2.5 times as high as the maximum stated rating is achieved. No thermal roll-over could be observed for even higher injection currents up to \SI{500}{mA}, which we address to a good thermal contact of the laser diode to the heatsink, resulting in an enhanced temperature stabilization of our compact setup. At \SI{500}{mA}, we measured an optical power of \SI{242}{mW}.

\label{subsec:pituning}

In the ECDL configuration the output power reaches \SI{65}{mW} at an injection current of \SI{200}{mA} and the lasing threshold is reduced from \SI{44}{mA} to \SI{36}{mA} compared to the free running configuration. We are not concerned about the back reflected power from the grating since it is not a major source of diode damage for GaN diodes, as in the case for GaAs diodes \cite{Tomm2017}.
The high output power of this ECDL is useful for a wide range of applications such as laser cooling and spectroscopy at cyan wavelengths around \SI{495}{nm}, where only SHG sources could deliver comparable power levels so far. The power could be further increased by injection locking a second diode in a master slave configuration. To further investigate the suitability of our laser source for atomic physics, wavelength tuning range and the laser linewidth are investigated.

\begin{figure}[htb]
\centering
\includegraphics[width=0.95\linewidth]{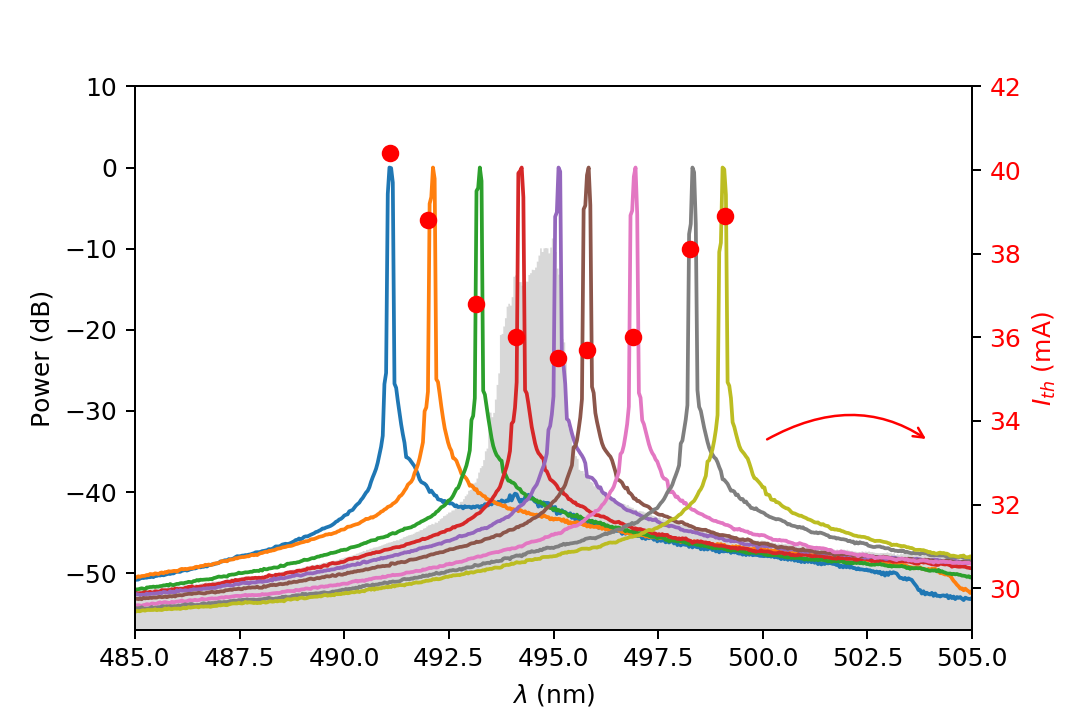}
\caption{Optical spectra of the ECDL for different wavelengths accessible with our configuration (colored) and for laser diode without feedback (grey) measured at \SI{25}{\celsius}. The red data points show the lasing threshold for each ECDL spectrum.}
\label{fig:tuning}
\end{figure}

The wavelength of the ECDL is tuned by changing the grating angle. Figure \ref{fig:tuning} shows the emission spectra of our ECDL for different wavelengths along with the emission spectrum of the bare diode (grey shadowed), measured with an optical spectrum analyzer (OSA). All measurements were performed at an injection current of \SI{105}{mA} and at a diode temperature of \SI{25}{\celsius}. A tuning range of \mbox{$\approx$\,\SI{8}{nm}} around the center wavelength, covering the range from \SI{491}{nm} to \SI{499}{nm} is achieved. This is comparable to commercial diode laser systems in the visible part of the spectrum and, to our knowledge, the first tunable diode laser covering this part of the spectrum. 

The resolution of the measurement is limited by our OSA to \SI{0.1}{nm}. The lasing threshold for each of the spectra is also shown in Figure \ref{fig:tuning}. It is strongly reduced compared to the bare diode and is the smallest for the wavelength of \SI{495}{nm}, the peak wavelength of the bare diode.

\subsection{Emission linewidth}
\label{subsec:Linewidth}

The power output of the ECDL is sufficient for laser spectroscopy and cooling applications with neutral strontium as has been shown above. Another criteria for the suitability of a laser in atomic physics is the linewidth. It should be narrower than the linewidth of the transition we want to address, in our case \SI{2.04}{MHz} for the $^3P_2 \rightarrow$  $^3D_2$ transition at \SI{496.93}{nm}.

\begin{figure}[htb]
\centering
\includegraphics[width=0.95\linewidth]{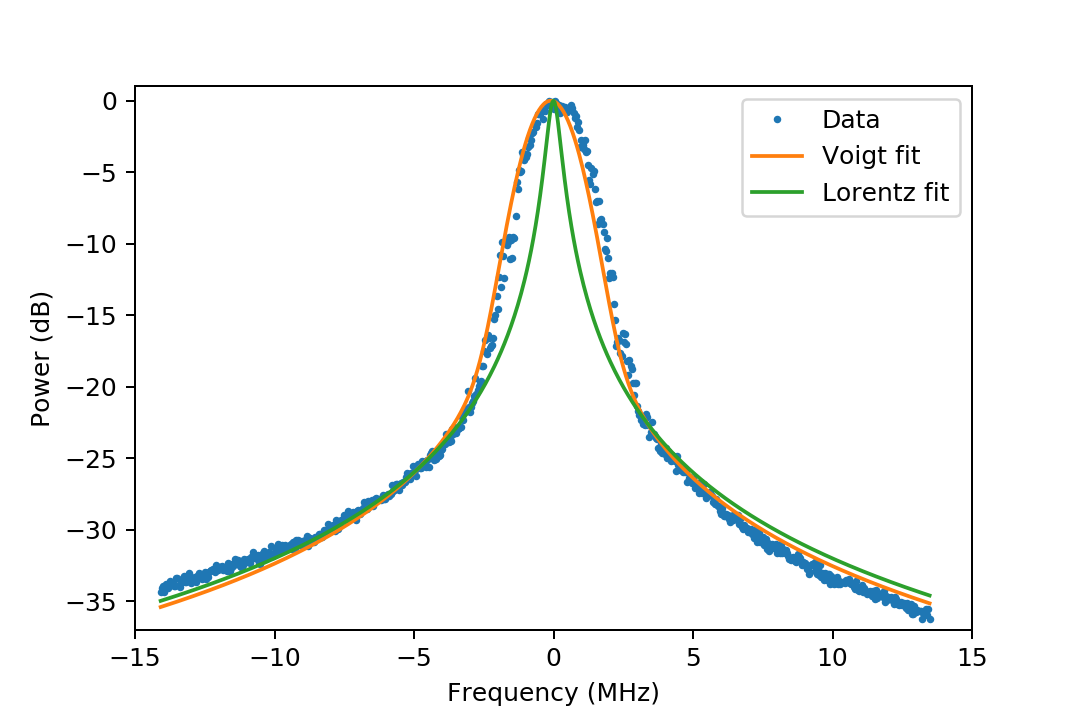}
\caption{Analysis of the beat note between two lasers for a sweep time of \SI{400}{ms} averaged for 35 sweeps. For a single laser this results in a Lorentzian linewidth of \SI{88}{kHz} and a Gaussian linewidth of \SI{1.22}{MHz}. See text for more details.}
\label{fig:beat}
\end{figure}

To investigate the linewidth of the diode laser we assembled a second laser and measured the free running beat note between both of them. \mbox{Figure \ref{fig:beat}} shows the analysis of our beatnote measurement where 35 sweeps were averaged for better statistics. Fitting a Voigt model to the data yields a Lorentzian linewidth of \SI{175 \pm 1}{kHz} and a Gaussian linewidth of \SI{1.72 \pm 0.01}{MHz} at a sweep time of \SI{400}{ms}. The Lorentzian linewidth is divided by 2 and the Gaussian linewidth is divided by $\sqrt{2}$ to obtain the respective linewidths
of a single laser. The resulting Lorentzian linewidth of \SI{88}{kHz} is comparable to commercial diode laser systems in the visible part of the spectrum. The Gaussian linewidth of \SI{1.22}{MHz} stems from external influences and from possible noise in our current driver and temperature controller. The linewidth could therefore be further reduced by low noise electronics and an improved laser housing that is less susceptible to mechanical and acoustic vibrations. Even without further improvement, the linewidth is already lower than the linewidths of the transitions in neutral strontium around \SI{497}{nm}.

\section{Conclusion}
\label{sec:conclusion}
We built and characterized an extended-cavity diode laser operating in the cyan wavelength range around \SI{497}{nm}. The laser provides an optical output power of \SI{65}{mW} at \SI{200}{mA} with only a slight deviation from a linear current/power relationship. A tuning range of $\approx$\,\SI{8}{nm} around the centre wavelength of \SI{495}{nm} was achieved. The linewidth of the laser was determined to be \SI{88}{kHz} for the Lorentzian part and \SI{1.22}{MHz} for the Gaussian part.\\
This compact, simple and low cost laser source can be used for several applications. Laser systems for cooling of strontium atoms can be considerably simplified in two different ways. First, the laser source can be utilized as a repump laser at \SI{496.93}{nm} to replace previous more complex systems and to reduce the number of needed laser sources. Second, sub-Doppler cooling at \SI{496.36}{nm} can be employed to replace the second cooling stage at \SI{689}{nm}. Furthermore these laser source facilitates quantum computing schemes using the $^3P_2 \rightarrow$ $^3D_3$ transition at \SI{496.36}{nm} for population detection \cite{Daley2011}, can be used for experimental investigation of the highly forbidden $5s\rightarrow6s$  transition in rubidium and potentially for compact optical frequency standards taking advantage of cyan transitions in molecular iodine \cite{Cheng2002}. Lastly, experiments employing cold barium ions can profit from convenient diode lasers for the cooling transition \mbox{$6s^2S_{1/2} \rightarrow 6p^2P_{1/2}$} at \SI{493.5}{nm} \cite{Hucul2017}.

\begin{acknowledgements}

This work is supported by the DLR Space Administration with funds provided by the Federal Ministry for Economic Affairs and Energy (BMWi) under grant numbers DLR 50WM1753 and 1857. 

The Authors would like to thank Achim Peters for useful discussions and providing resources, Klaus D{\"{o}}ringshoff for useful discussions and proofreading the manuscript.
\end{acknowledgements}



\bibliographystyle{ieeetr}
\bibliography{Schkolnik_Sr_Ls}

\end{document}